\documentclass
{aa}

\usepackage{graphicx}
\usepackage{natbib}
\usepackage{txfonts}
\bibpunct{(}{)}{;}{a}{}{,}

\def\3c{3C\,273}
\def\PI{Paper~I}
\def\Lya{Ly$\alpha$}
\def\Lyal{Ly$\alpha$ $\lambda$1216}
\def\Lnnv{Ly$\alpha-$N{\sc v}}
\def\Lynv{Ly$\alpha+$N{\sc v}}
\def\nv{N{\sc v}}
\def\nvl{N{\sc v} $\lambda$1240}
\def\civ{C{\sc iv}}
\def\civl{C{\sc iv} $\lambda$1549}
\def\Ha{H$\alpha$}
\def\Hb{H$\beta$}
\def\Hg{H$\gamma$}
\def\kms{\mbox{km\,s$^{-1}$}}

\def\dd{{\mathrm{d}}}
\def\un{{\phantom{1}}}

\begin{document}
\title{The mass of the black hole in \3c}

\author{S. Paltani\inst{1}\thanks{\emph{Present address:} INTEGRAL
    Science Data Centre, ch.\ d'\'Ecogia 16, CH-1290 Versoix,
    Switzerland} \and M. T\"urler\inst{2,3}}

\institute{Laboratoire d'Astrophysique de Marseille, BP 8, 13376 Marseille cedex 12, France 
  \and INTEGRAL Science Data Centre, ch.\ d'\'Ecogia 16, CH-1290 Versoix, Switzerland
  \and Geneva Observatory, ch. des Maillettes 51, CH-1290 Sauverny, Switzerland}

\offprints{S. Paltani, \email{Stephane.Paltani@ obs.unige.ch}}

\date{Received 3 May 2004 / Accepted 21 January 2005}

\abstract{In this paper we apply the reverberation method to determine
  the mass of the black hole in \3c\ from the \Lya\ and \civ\ emission
  lines using archival IUE observations.  Following the standard
  assumptions of the method, we find a maximum-likelihood estimate of
  the mass of $6.59\,10^9$\,M$_\odot$, with a $1\sigma$ confidence
  interval $5.69-8.27\,10^9$\,M$_\odot$. This estimate is more than
  one order of magnitude larger than that obtained in a previous study
  using Balmer lines. We reanalyze the optical data and show that the
  method applied to the \Ha, \Hb, and \Hg\ Balmer lines produce mass
  estimates lower by a factor 2.5, but already much larger than the
  previous estimate derived from the same lines. The finding of such a
  high mass in a face-on object is a strong indication that the gas
  motion is not confined to the accretion disk. The new mass estimate
  makes \3c\ accreting with an accretion rate about six times lower
  than the Eddington rate.  We discuss the implications of our result
  for the broad-line-region size and black-hole mass vs luminosity
  relationships for the set of objects for which reverberation
  black-hole masses have been obtained. We find that, while objects
  with super-Eddington luminosities might theoretically be possible,
  their existence is not necessarily implied by this sample.
  
   }

\maketitle

\section{Introduction}
\label{sec:intro}
As soon as the variability of the broad emission lines in Seyfert 1
galaxies and QSOs had been discovered, its potential for the
investigation of the structure of these objects became evident.
\citet{GaskSpar-1986-LinVar} were the first to propose using
cross-correlation methods between the continuum and line light curves
to measure the size of the broad-line region (BLR).  The subsequent
building of long and dense spectro-photometric light curves of
broad-line active galactic nuclei (BL-AGN), in particular thanks to
the IUE satellite and to the AGN Watch consortium
\citep{AlloEtal-1994-IntAgn}, made the reliable measurement of the BLR
size in a significant number of objects possible. The so-called
reverberation method was later developed to allow a determination of
the mass of the central black hole
\citep{PeteWand-1999-KepMot,WandEtal-1999-CenMas}, under the
assumption that the gas dynamics is dominated by gravity and using the
fact that its velocity dispersion can be deduced from the width of the
line.  The connection between black-hole mass and line widths in
BL-AGN has allowed the development of methods to determine the
black-hole masses in high-redshift QSOs
\citep{Vest-2002-DetCen,McluJarv-2002-MeaBla}, and thus opened the
door to cosmological studies of black holes. These methods require,
however, a good understanding of the relationship between luminosity
and BLR size, which must be investigated through the use of the
reverberation method.

\citet{Krol-2001-SysErr} warned that some assumptions made in
the context of the reverberation method are not sufficiently
validated. A key assumption is that the gas dynamics is dominated by
the gravitational pull from the central black hole. While
\citet{PeteWand-1999-KepMot} showed that the $v\propto r^{-1/2}$
relationship expected in such case is satisfied in \object{NGC\,5548}
for different lines, other competing models can also explain this
dependence.  The best support for the assumptions behind the
reverberation method is the agreement between the reverberation masses
and those obtained using stellar-velocity dispersion
\citep{GebhEtal-2000-BlaHol}.  \citet{MerrFerr-2001-RelSup} also
concluded that the reverberation-mass estimates are not more strongly
biased than stellar velocity-dispersion ones.  The existence of
hidden assumptions nevertheless makes \citet{Krol-2001-SysErr} conclude
that, even if valid, the method is subject to systematic
uncertainties as large as a factor 3 or more.

In a previous paper \citep[][hereafter \PI]{PaltTurl-2003-DynLya}, we
studied in detail the response of the \Lya\ and \civ\ emission lines
in \object{\3c}.  In this paper, we extend our study by applying the
reverberation method to determine the mass of the central black hole
in this object. We make the assumption that the reverberation method
is valid and apply the methodology presented in
\citet{WandEtal-1999-CenMas} to derive the black-hole mass using the
strongest broad ultraviolet lines: \Lya\ and \civ. We differ with
previous works on the treatment of statistical errors and
propose here a full error propagation, in order to obtain meaningful
confidence intervals on our results. We also compare our results with
those in the optical domain from \citet{KaspEtal-2000-RevMea}. We
finally assess the consequence of our results on two important
relationships in BL-AGN: the broad-line-region size vs luminosity
and black-hole mass vs luminosity relationships.

\section{Data}
\label{sec:data}
This paper makes use of data from the \3c\ database hosted by the
Geneva Observatory\footnote{Web site:
  \tt{http://obswww.unige.ch/3c273/}} and presented in
\citet{TuerEtal-1999-ThiYea}. We use the same data set as in \PI,
namely the IUE short wavelength (SWP) spectra processed with the IUE
Newly Extracted Spectra (INES) software \citep{RodrEtal-1999-IueIne}.
Details of data preparation can be found in \PI, but we summarize
below the main properties of the sample for completeness.

Our light curves each consist of 119 observations covering 18 years,
including a ten-year period with semi-weekly observations during two
annual observation periods of 3 months, separated by two months. From
the original 256 SWP observations, over- and underexposed observations
were discarded, and observations taken within 12 hours were averaged.
The \Lyal\ and \civl\ fluxes were integrated above a continuum defined
by a straight line fitted to the points in two 50\,\AA\ continuum
bands on each side of the lines.  We integrate the total \Lya\ and
\civ\ line fluxes for projected rest-frame velocities up to $\pm
20\,000$\,\kms\ and $\pm 10\,000$\,\kms\ respectively. Uncertainties
on the line fluxes take into account the uncertainty in the underlying
continuum and the number of independent bins. The \Lya\ emission line
is contaminated by \nvl\ emission. In the following, ``\Lynv'' will be
used to designate the full line profile around \Lya, which includes
the \nv\ contamination, and ``\Lnnv'' to designate the line profile
around \Lya\ where the range 3500--10\,000\,\kms\ has been linearly
interpolated to get rid of the \nv\ contamination.

\begin{figure}[tbp]
  \resizebox{\hsize}{!}{\includegraphics{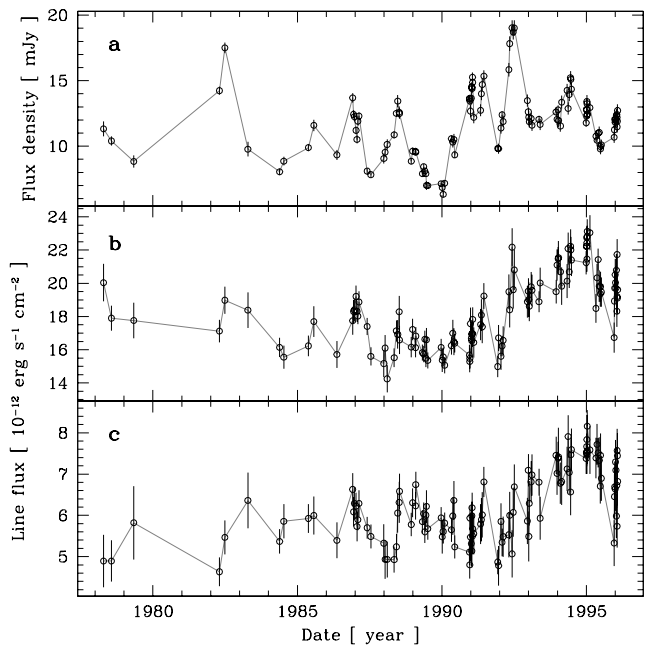}}%
  \caption{\label{fig:lcurve}
    \textbf{(a)} Ultraviolet continuum light curve averaged in the
    1250--1300\,\AA\ band.  \textbf{(b)} \Lynv\ emission line light curve
    integrated for velocities up to 20\,000\,\kms. \textbf{(c)} \civ\ 
    emission line light curve integrated for velocities up to
    10\,000\,\kms.}
\end{figure}

Figure~\ref{fig:lcurve} shows the three main light curves used in this
study: the 1250--1300\,\AA\ continuum, the integrated \Lynv\ emission
line, and the integrated \civ\ emission line. All light curves used in
this paper are listed in Table 1 of \PI, which is available
electronically.  They can also be downloaded from the \3c\ database.

\section{Black-hole mass}
We estimate the mass of the black hole in \3c\ using the reverberation
method introduced by \citet{PeteWand-1999-KepMot}. The line-emitting
gas is assumed to be in virial equilibrium around a mass dominated by
the black hole, so that the gas distance to the central object and
its velocity dispersion are both related to the mass through the virial
theorem. The velocity dispersion can be estimated from the line width,
and the gas distance from the delay between the continuum and line
light curves, making possible a determination of the mass. In
principle, a single line is sufficient.  The consistency of mass
estimates from different lines is, however, a stringent test of the
validity of the method.

In \PI, we showed that the \Lya\ and \civ\ emission lines show two
distinct components.  We interpreted the second component as evidence
of gas entrainment in the course of an interaction with the jet.  In
principle, the presence of a non-virial velocity field makes the above
method inapplicable; however, we showed that this component mostly
affects the blue wings of the lines, making the red sides (hereafter,
the ``red profiles'') suitable for a mass determination.  We
nevertheless perform the analysis both on the full lines and on the
red profiles for comparison. We also perform the analysis separately
on \Lynv\ and \Lnnv\ to evaluate the effect of contamination by \nv.

In this work, we shall make sure to correctly propagate all
uncertainties in our data into the determination of the mass. As the
error distributions of the lag and velocity dispersion are not at all
Gaussian, they cannot be combined analytically to determine the
confidence intervals on the black-hole mass. We thus construct the
full black-hole mass error distribution using those of the lag and of
the velocity dispersion.

\subsection{Distance of the line-emitting gas}
The distance of the line-emitting gas to the source can be estimated
using the delay between the line and ionizing-continuum (or its
closest approximation) light curves.  Delays can be readily obtained
using a cross-correlation analysis.  However, as the gas fills a vast
range of distances, there is no unique way to derive a delay from a
cross-correlation.  \citet{KoraGask-1991-StrKin} have argued that the
centroid of the cross-correlation function gives the
emissivity-weighted average radius of the emission region.

\begin{figure}[tbp]
  \resizebox{\hsize}{!}{\includegraphics{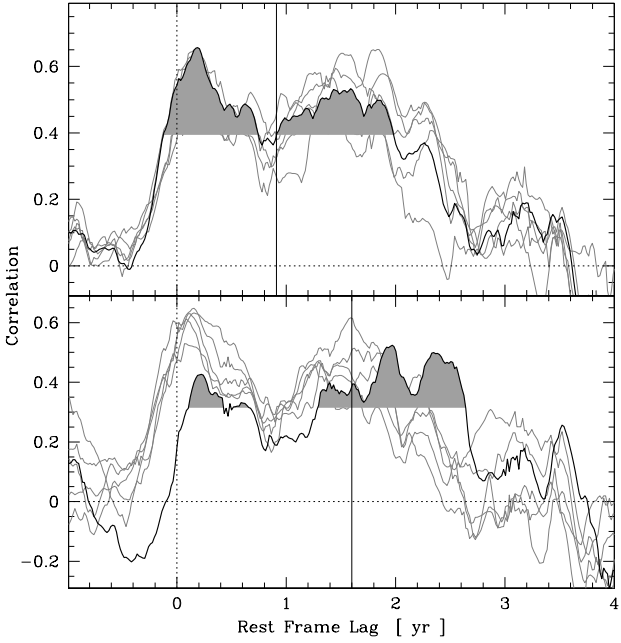}}%
  \caption{\label{fig:cross-peak}
    Cross-correlation between the UV continuum and the red-profile
    \Lynv\ (top) and \civ\ (bottom) light curves. In both panels, the
    shaded area represents the parts of the correlation above the
    threshold set at 60\% of its maximum (i.e.  $\vartheta=0.6$). Only
    these areas are used in the estimation.  The average lag is
    indicated by a vertical line. Five realizations of the FR/RSS
    method are also shown in each panel for illustration.}
\end{figure}

We follow here the method proposed by \citet{PeteEtal-1998-UncCro}. We
calculate the cross-correlations between the continuum (in the
1250--1300\,\AA\ band) and the \Lynv, \Lnnv, and \civ\ light curves.
We use the Interpolated Cross-Correlation Function (ICCF) of
\citet{GaskPete-1987-AccCro} with the improvements introduced by
\citet{WhitPete-1994-ComCro}. The delay is estimated by calculating
the average lag weighted by the correlation coefficient for all those
parts of the cross-correlation which exceed a given threshold
$\vartheta$ expressed as a fraction of the maximum correlation.  This
is illustrated in Fig.~\ref{fig:cross-peak} for the actual \Lynv\ and
\civ\ correlations.
The lag distribution is estimated using the
flux-randomization/random-subset-selection (FR/RSS) method of
\citet{PeteEtal-1998-UncCro}: Both line and continuum fluxes are first
randomized according to their uncertainties, and 119 dates are drawn
at random (with possible repetitions) from the original dates.  The
two light curves are then cross-correlated, and a new lag estimate is
produced. The lag error distribution is then constructed from the
distribution of lag estimates. Examples of FR/RSS realizations
are shown in Fig.~\ref{fig:cross-peak}.

\begin{figure}[tbp]
  \resizebox{\hsize}{!}{\includegraphics{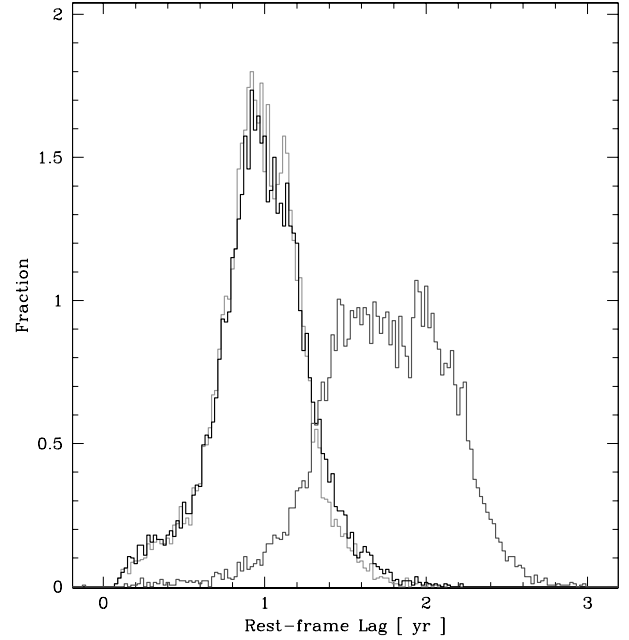}}%
  \caption{\label{fig:laghst}
    Lag error distribution for the red profiles of the \Lnnv\ (black),
    \Lynv\ (light grey), and \civ\ (dark grey) emission lines}
\end{figure}

Because one wants to take into account as much gas as possible, the
threshold should be set as low as possible.  However, if taken too
low, non-significant parts of the correlation function are included
in the lag estimate. Considering the flat-top correlation functions we
observe here, high thresholds seem inappropriate, as they make the
result quite dependent on small details of the correlation functions.
We adopt here a threshold of 0.6, but we checked that the choice of
the threshold has a moderate impact of about 30\% on the average delay
throughout the range $0.3\le\vartheta\le 1$. The error distributions,
however, become very large and are no longer unimodal as soon as
$\vartheta>0.8$, which shows that the lags start being dominated by
non-significant details of the correlation function. The average lags
and the confidence intervals are given in Table~\ref{tab:lag}. All
lags are expressed in the object's rest frame. In both \Lya\ and \civ\ 
lines, the lags of the full profiles are systematically larger than
their corresponding lags in the red profiles, consistent with the
results of \PI. The individual differences are not statistically
significant however.  The effect of \nv\ is completely negligible.
Error distributions of the lags for the red profiles are shown on
Fig.~\ref{fig:laghst}.
\begin{table}[tbp]
  \caption{\label{tab:lag}
    Average rest-frame lags in years of the \Lya\ and \civ\ emission
    lines. The numbers in brackets give the $\pm 1\sigma$, $\pm 2\sigma$,
    and $\pm 3\sigma$ confidence intervals.}
  \begin{center}
    \begin{tabular}{lcc}
      \hline
      \hline
      \rule{0pt}{1.2em}Line&Full profile&Red profile\\
      \hline\rule{0pt}{1.2em}%
      \rule[-0.6em]{0pt}{1em}\Lynv&$1.19~\left\{_{-0.21}^{+0.21}\right.\left\{_{-0.46}^{+0.50}\right.\left\{_{-0.86}^{+0.79}\right.$&
      $0.97~\left\{_{-0.24}^{+0.25}\right.\left\{_{-0.63}^{+0.54}\right.\left\{_{-0.85}^{+0.92}\right.$\\
      \rule[-0.6em]{0pt}{1em}\Lnnv&$1.21~\left\{_{-0.20}^{+0.20}\right.\left\{_{-0.44}^{+0.49}\right.\left\{_{-0.82}^{+0.79}\right.$&
      $0.98~\left\{_{-0.26}^{+0.27}\right.\left\{_{-0.68}^{+0.60}\right.\left\{_{-0.86}^{+1.00}\right.$\\
      \rule[-0.6em]{0pt}{1em}\civ&$1.89~\left\{_{-0.29}^{+0.28}\right.\left\{_{-0.53}^{+0.55}\right.\left\{_{-0.89}^{+0.80}\right.$&
      $1.75~\left\{_{-0.37}^{+0.39}\right.\left\{_{-0.80}^{+0.69}\right.\left\{_{-1.49}^{+1.08}\right.$\\
      \hline
    \end{tabular}
  \end{center}
\end{table}

\subsection{Line velocity dispersion}
The velocity dispersion of the emitting gas can be estimated using the
shape of the line.  In the context of mass determination, line
velocity dispersion is generally determined from the full width at
half maximum (FWHM) of the rms profile \citep{WandEtal-1999-CenMas}.
This profile is then used to avoid contamination from a possible
stable narrow-line component.  The relationship between FWHM and gas
velocity dispersion relies, however, on assumptions about the line
shape that are difficult to assess. In addition, the measurement of
the FWHM is very sensitive to details of the profile.
\citet{FromMeli-2000-DetCen} propose to calculate the squared velocity
dispersion directly from the second moment of the line profile:
\begin{equation}
<\!v^2\!> = 3 \left( \frac{\int P(v) v^2 \dd v}{\int P(v) \dd v}-\left(\frac{\int P(v) v \dd v}{\int P(v) \dd v}\right)^2 \right),
\label{eq:disp}
\end{equation}
where $P(v)$ is the amplitude of the rms profile at rest-frame
velocity $v$.  The only assumption behind this method is the isotropy
of the emission, which is the origin of the factor 3 in the above
equation.

\begin{figure}[tbp]
  \resizebox{\hsize}{!}{\includegraphics{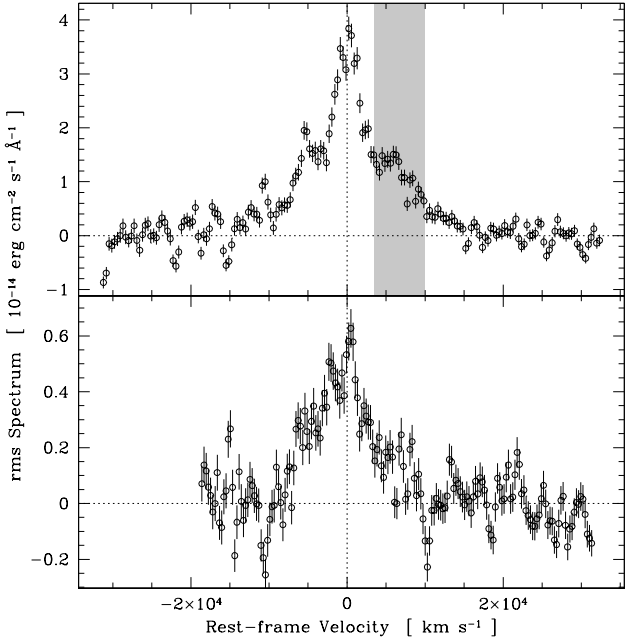}}%
  \caption{\label{fig:fwhmrms}
    Continuum-subtracted rms profiles of the \Lynv\ (top) and \civ\ 
    (bottom) emission lines. The grey area marks the region of the
    \Lya\ profile contaminated by \nv\ emission line.}
\end{figure}

Figure~\ref{fig:fwhmrms} shows the two rms profiles of the \Lynv\ 
and \civ\ emission lines.
The rms profiles are rather complex, presenting a narrow core, broad
wings and secondary peaks, and are not easily represented by a simple
analytic form.  This is the kind of situation to which the method of
\citet{FromMeli-2000-DetCen} is particularly adapted; in the
following, we adopt Eq.~(\ref{eq:disp}) to calculate the velocity
dispersion.
  
It is not clear how the error distribution of the velocity dispersion
can be calculated. \citet{KaspEtal-2000-RevMea} use the FWHM
dispersion measured in individual spectra as an estimate of the
uncertainty in the FWHM of the rms profile, but there is no obvious
reason why it would give the correct result.  Here we follow a
different approach: The uncertainty in any point of the rms profile
can be estimated using the standard estimator of error on the rms:
$\sigma_{\mathrm{rms}}(\lambda)=\mathrm{rms}(\lambda)/\sqrt{2(N-1)}$,
where $N$ is the number of spectra.  However, the rms in different
wavelength bins will be strongly correlated with each other, because
they have been obtained using the same spectra.  Thus we scale the
uncertainties on the rms profile by requiring that the average
uncertainty of the rms profile in the continuum regions is equal to
the standard deviation of the rms estimates.  We can then estimate the
velocity-dispersion distribution by randomizing the rms profiles a
large number of times according to the uncertainties in each bin
before continuum subtraction (so that the error on the continuum
determination is included in our determination), and by measuring the
velocity dispersion using Eq.~(\ref{eq:disp}) for each of them.

\begin{figure}[tbp]
  \resizebox{\hsize}{!}{\includegraphics{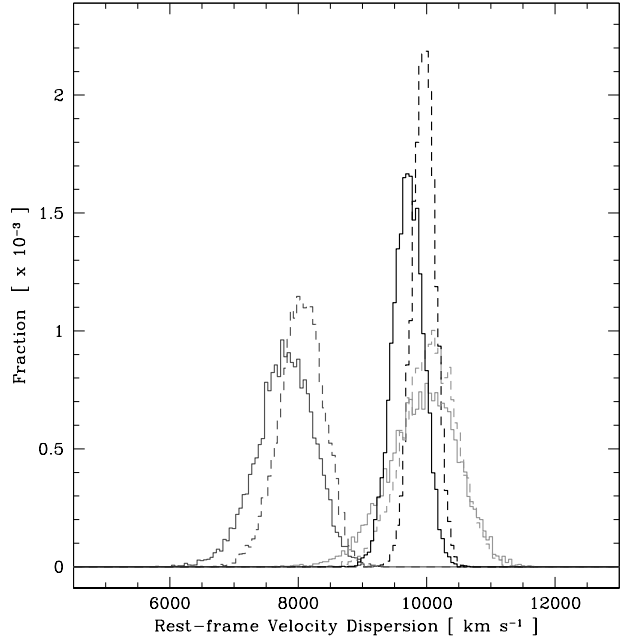}}%
  \caption{\label{fig:fwhmhst}
    Velocity-dispersion error distributions of the red profiles
    (continuous lines) and full profiles (dashed lines) of the \Lnnv\ 
    (black), \Lynv\ (light grey), and \civ\ (dark grey) rms profiles.
  }
\end{figure}

Table~\ref{tab:fwhm} gives the velocity dispersions of the \Lya\ and
\civ\ emission lines obtained with the 2nd-moment method.
Figure~\ref{fig:fwhmhst} compares the velocity-dispersion error
distribution for the red profiles of \Lnnv\ and \civ. We removed a
spectral bin close to the center of the \civ\ line with a linear
interpolation, because it was strongly affected by a reseau mark.
\begin{table}[tbp]
  \caption{\label{tab:fwhm}
    Average velocity dispersions in \kms\ of the \Lya\ and \civ\
    emission lines.  The numbers in brackets give the $\pm 1\sigma$,
    $\pm 2\sigma$, and $\pm 3\sigma$ confidence intervals.}
  \begin{center}
    \begin{tabular}{l@{~~~}c@{~~~}c}
      \hline
      \hline
      \rule{0pt}{1.2em}Line&Full profile&Red profile\\
      \hline\rule[-0.6em]{0pt}{1.8em}\Lynv&$10\,044~\left\{_{-423}^{+423}\right.\left\{_{-892}^{+812}\right.\left\{_{-1341}^{+1119}\right.$&
$\un 9\,941~\left\{_{-539}^{+532}\right.\left\{_{-1133}^{+\un 996}\right.\left\{_{-1909}^{+1431}\right.$\\
      \rule[-0.6em]{0pt}{1em}\Lnnv&$\un 9\,955~\left\{_{-178}^{+179}\right.\left\{_{-366}^{+347}\right.\left\{_{-\un 575}^{+\un 512}\right.$&
      $\un 9\,707~\left\{_{-240}^{+240}\right.\left\{_{-\un 491}^{+\un 461}\right.\left\{_{-\un 738}^{+\un 674}\right.$\\
      \rule[-0.6em]{0pt}{1em}\civ&$\un 8\,049~\left\{_{-346}^{+352}\right.\left\{_{-738}^{+667}\right.\left\{_{-1129}^{+1003}\right.$&
      $\un 7\,795~\left\{_{-436}^{+442}\right.\left\{_{-\un 947}^{+\un 828}\right.\left\{_{-1557}^{+1137}\right.$\\
      \hline
    \end{tabular}
  \end{center}
\end{table}
The effect of removing \nv\ is sufficiently small so that we can
exclude a strong residual \nv\ contamination in \Lnnv.  The effect of
asymmetry on the measurement of the velocity dispersion is
approximately 3\% in both \Lnnv\ and \civ. This consistency shows
that the line is dominated by a symmetric velocity component, assumed
to be due to Keplerian motion, but the broadening of the full profile
with respect to the red profile, although not significant, is
systematic.

\subsection{Mass determination}
\label{sec:ML} 
The velocity dispersion and delay can be combined to obtain the
black hole mass from the virial equation:
\begin{equation}
  \label{eq:mass}
  M_{\mathrm{BH}}=71.31~\tau <\!v^2\!> \mathrm{M}_\odot
\end{equation}
where $\tau$ is the lag expressed in units of years and
$<\!\!v^2\!\!>$ the squared velocity dispersion expressed in units of
km s$^{-1}$. This formula is identical to Eq.~(6) of
\citet{WandEtal-1999-CenMas}, except that the velocity dispersion is
used instead of the FWHM. As in \citet{PeteWand-1999-KepMot}, the
factor between the kinetic and potential energies in the virial
theorem is set more or less arbitrarily to $-1$. Error
distributions of the masses are constructed using a Monte-Carlo
approach. We draw a large number of pairs of random values from the
error distributions of the lags and velocity dispersions, and combine
them using Eq.~(\ref{eq:mass}) to obtain the error distributions of
the masses.  The results are given in Table~\ref{tab:mass}.
\begin{table}[tbp]
  \caption{\label{tab:mass}
    Average black-hole masses of \3c\ in units of $10^9\,$M$_\odot$.
    The numbers in brackets give the $\pm 1\sigma$,
    $\pm 2\sigma$, and $\pm 3\sigma$ confidence intervals. $\hat{M}$
    is the maximum-likelihood mass obtained from the combination of
     \Lnnv\ and \civ\ emission lines.}
  \begin{center}
    \begin{tabular}{lcc}
      \hline
      \hline
      \rule{0pt}{1.2em}Line&Full profile&Red profile\\
      \hline
      \rule[-0.6em]{0pt}{1.8em}\Lynv&$8.56~\left\{_{-1.68}^{+1.72}\right.\left\{_{-3.50}^{+4.01}\right.\left\{_{-6.04}^{+6.40}\right.$&
      $6.83~\left\{_{-1.88}^{+1.91}\right.\left\{_{-4.52}^{+4.25}\right.\left\{_{-6.02}^{+7.10}\right.$\\
      \rule[-0.6em]{0pt}{1em}\Lnnv&$8.56~\left\{_{-1.40}^{+1.47}\right.\left\{_{-3.19}^{+3.51}\right.\left\{_{-5.74}^{+5.91}\right.$&
      $6.59~\left\{_{-1.77}^{+1.83}\right.\left\{_{-4.55}^{+4.07}\right.\left\{_{-5.79}^{+6.67}\right.$\\
      \rule[-0.6em]{0pt}{1em}\civ&$8.75~\left\{_{-1.53}^{+1.53}\right.\left\{_{-2.79}^{+3.15}\right.\left\{_{-4.41}^{+4.90}\right.$&
      $7.62~\left\{_{-1.80}^{+1.90}\right.\left\{_{-3.64}^{+3.77}\right.\left\{_{-6.50}^{+6.20}\right.$\\
      \hline
\rule[-0.6em]{0pt}{1em}\rule{0pt}{1.2em}$\hat{M}$&
      $8.20~\left\{_{-1.14}^{+1.02}\right.\left\{_{-2.25}^{+2.31}\right.\left\{_{-3.16}^{+3.82}\right.$&
      $6.59~\left\{_{-0.90}^{+1.68}\right.\left\{_{-1.87}^{+3.04}\right.\left\{_{-3.15}^{+4.62}\right.$\\
      \hline
    \end{tabular}
  \end{center}
\end{table}
\begin{figure}[tbp]
  \resizebox{\hsize}{!}{\includegraphics{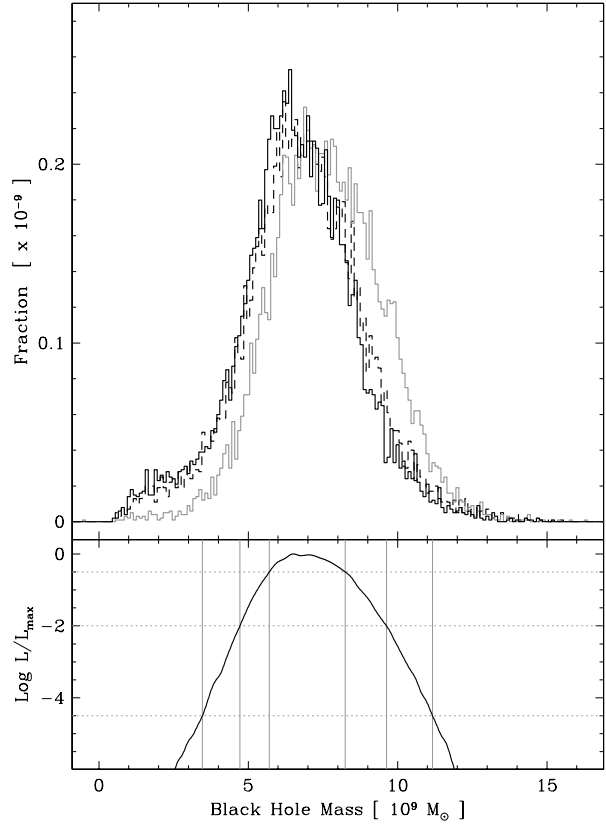}}%
  \caption{\label{fig:mass}
    Top panel: Black-hole mass distributions for the red profiles of
    the \Lnnv\ (black), \Lynv\ (dashed line), and \civ\ (grey)
    emission lines. Bottom panel: Likelihood function of the mass.
    The horizontal grey lines show, from top to bottom, the $\pm
    1\sigma$, $\pm 2\sigma$, and $\pm 3\sigma$ levels. The vertical
    lines show the $\hat{M}_{\pm 1}$, $\hat{M}_{\pm 2}$, and
    $\hat{M}_{\pm 3}$ intervals.}
\end{figure}
As expected from Paper~I, the average masses obtained with the red
profiles of the lines are systematically smaller than those obtained
with the full profiles. The full-profile values fall however within,
or close to, the $\pm 1\sigma$ red-profile confidence interval.  The
difference between \Lynv\ and \Lnnv\ is less than 5\% for the red
profiles.  The minimum mass at the $3\sigma$ level is about
$0.8\,10^9$ M$_\odot$.

The black-hole mass distributions for the red profiles of the \Lynv,
\Lnnv, and \civ\ emission lines are shown in Fig.~\ref{fig:mass}. All
three distributions overlap very significantly.  Confronted with
several estimates of the same values, we can combine them using a
$\chi^2$ test only if the error on the estimates have Gaussian
distributions.  As this is not the case here, we use a
maximum-likelihood estimator (MLE), where mass is the unknown
parameter that we try to recover.  In the simple case of $N$
independent measurements of the same value that we consider here, the
(natural) logarithm of the likelihood function is a function of the
mass $M$ given by:
\begin{equation}
\log L(M)=\sum_{i=1}^{N} \log P_i(M),
\end{equation}
where $P_i(M)$ is the probability distribution of $M$ in measurement
$i$ and is approximated by our black-hole mass distributions. The MLE
$\hat{M}$ is then the value of the mass that maximizes $L(M)$:
\begin{equation}
\frac{\dd \log L(\hat{M})}{\dd M}=0.
\end{equation}
The $\{\hat{M}_{-n},\hat{M}_{+n}\}$ $n$-sigma confidence interval can
be determined using:
\begin{equation}
\log L(\hat{M}_{\pm n})=\log L(\hat{M})-\frac{n^2}{2}.
\end{equation}

Applying this method to measurements of the black hole mass for the
red profiles of the \Lnnv\ and \civ\ emission lines, we obtain a
maximum-likelihood mass of $6.59\,10^9$ M$_\odot$. The $1\sigma$,
$2\sigma$, and $3\sigma$ confidence intervals are $5.69-8.27\,10^9$
M$_\odot$, $4.72-9.63\,10^9$ M$_\odot$, and $3.44-11.21\,10^9$
M$_\odot$ respectively. The difference in the MLEs for the full and
red profiles is about 1.5$\sigma$.

\section{Mass derived from Balmer lines}
\citet{KaspEtal-2000-RevMea} (hereafter K00) performed a series of
spectro-photometric observations in the 4500--8000\,\AA\ domain on a
set of 28 QSOs, among which is \3c. From the Balmer lines, they
obtained a mass of $0.235^{+0.037}_{-0.033}\,10^9$ M$_\odot$ from the
rms profiles, and $0.550^{+0.089}_{-0.079}\,10^9$ M$_\odot$ from the
average line profiles.  These values are apparently the average of the
masses obtained using \Ha\ and \Hb. Individual estimates of the mass
for different lines and different line-width determination methods
vary by a factor 5 from $0.115$ to $0.56\,10^9$ M$_\odot$, i.e.\ a
much larger range than the $\sim\!15$\% uncertainties they claim.
This indicates a serious problem in their mass determination for this
object. These masses are in complete disagreement with our result
using \Lya\ and \civ, as they are between 12 and 60 times smaller than
our maximum-likelihood determination, and at least $6$ times smaller
than $\hat{M}_{-3}$.

\begin{figure}[tbp]
  \resizebox{\hsize}{!}{\includegraphics{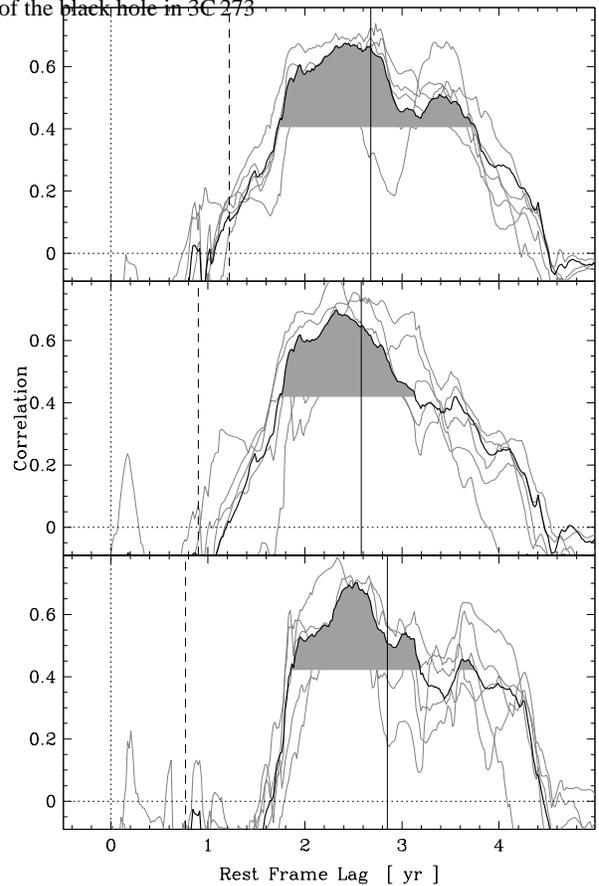}}%
  \caption{\label{fig:crossbalm-peak}
    Cross-correlation between the UV continuum and the \Ha\ (top),
    \Hb\ (center), and \Hg\ (bottom) light curves. In all three
    panels, the shaded area represents the parts of the correlation
    above the threshold set at 60\% of its maximum (i.e.,
    $\vartheta=0.6$). Only these areas are used in the weighted
    average of the lag (indicated by a vertical line). The dashed
    vertical lines indicate the lags from K00. Five realizations of
    the FR/RSS method are also shown in each panel for illustration. }
\end{figure}

It must first be noted that K00 determined the line lag with respect
to continuum using the optical continuum around 5000\,\AA.
\citet{PaltEtal-1998-BluBum} showed that the optical continuum
properties of \3c\ differ considerably from those of the ultraviolet
continuum.  They concluded that the optical emission of \3c\ is
strongly contaminated, and perhaps even dominated, by non-thermal
radiation, possibly related to the jet of \3c, and varying on much
longer time scales than the UV continuum.  The optical continuum
therefore appears unsuitable for studying the lag between the ionizing
continuum and the lines. Using the (full-line) \Ha, \Hb\ and \Hg\ 
light curves from K00, we redetermine the lags with the same
ultraviolet continuum light curve used to determine the lags with
\Lya\ and \civ. In addition to using a continuum much closer to the
ionizing continuum, the lag determination is improved by mutiplying by
three the number of observations over a twice as long period of time.
Fig.~\ref{fig:crossbalm-peak} shows the correlation between the UV
continuum and the three Balmer lines \Ha, \Hb, and \Hg.
Table~\ref{tab:lag_balmer} compares our lags with those from K00, ours
being between 2 and 4 times larger.  Furthermore, our estimates are in
excellent agreement with each other, while in K00 the \Ha\ lag is 60\%
larger than that of \Hg.
\begin{table}[tbp]
  \caption{\label{tab:lag_balmer}
    Average rest-frame lags in years of the \Ha, \Hb, and \Hg\ emission
    lines found in this study compared to the values of K00. The numbers
    in brackets give the $\pm 1\sigma$, $\pm 2\sigma$, and $\pm 3\sigma$
    confidence intervals. Only the $\pm 1\sigma$ confidence intervals are
    available from K00.}
  \begin{center}
    \begin{tabular}{lcc}
      \hline
      \hline
      \rule{0pt}{1.2em}Line&UV-continuum lag&Optical-continuum lag\\
      &(This work)&(K00)\\
      \hline\rule{0pt}{1.2em}%
      \rule[-0.6em]{0pt}{1em}\Ha&$2.68~\left\{_{-0.15}^{+0.14}\right.\left\{_{-0.36}^{+0.33}\right.\left\{_{-0.56}^{+0.58}\right.$&
      $1.22~\left\{_{-0.16}^{+0.15}\right.$\\
      \rule[-0.6em]{0pt}{1em}\Hb&$2.58~\left\{_{-0.19}^{+0.19}\right.\left\{_{-0.33}^{+0.41}\right.\left\{_{-1.20}^{+0.71}\right.$&
      $0.90~\left\{_{-0.22}^{+0.28}\right.$\\
      \rule[-0.6em]{0pt}{1em}\Hg&$2.85~\left\{_{-0.32}^{+0.32}\right.\left\{_{-0.69}^{+0.71}\right.\left\{_{-2.63}^{+1.00}\right.$&
      $0.77~\left\{_{-0.25}^{+0.14}\right.$\\
      \hline
    \end{tabular}
  \end{center}
\end{table}

The velocity dispersion measurements in K00 are also problematic for
\3c.  The rms widths of the Balmer lines in \3c\ reported by K00 are
completely incompatible with each other, differing by a factor 2.5,
while the quoted uncertainties on the widths are only about 2--6\%.
Such disagreements are absent in most objects of their sample. As we
do not have access to the spectra used in K00, we use two
spectro-photometric observations performed with the STIS instrument
(\texttt{O44301010} for \Hg\ and \texttt{O44301020} for \Hb\ and \Ha)
on the Hubble Space Telescope.  Without variability information, we
have to apply the second-moment method to the average profile
(calculated with a single profile) instead of the rms profile. The
continuum and line regions were defined exactly as in K00.  In theory,
the line width can be reduced by the presence of narrow-line Balmer
emission, but K00 showed that the effect over their whole sample is
small.  Because of strong [O\,{\sc iii}] 4959\AA\ and 5007\AA\ 
contamination in the red profile of \Hb, we used only the blue profile
of this line.  Some contaminating emission lines are present in the
\Ha\ and \Hg\ profiles, but we checked that the blue and red profiles
give results in good agreement, showing that the contamination is
probably small.  The errors on these values were arbitrarily set to
10\%, because, if calculated as in the case of the UV lines, the
formal errors would be unrealistically low.
Table~\ref{tab:fwhm_balmer} compares our velocity dispersions with
those in K00. Except for \Ha, which is 80\% larger, the widths
obtained here are comparable to those in K00 using the mean profile.
The rms velocity-dispersion measurements of K00 for \Hb\ and \Hg\ are
about 15--25\% smaller, and the discrepancy reaches a factor 4 for
\Ha. Considering that the K00 values for the different Balmer lines
using the mean profile differ by 25\% and those using the rms profile
by a factor 2.5, it seems to us that there could be a problem in their
rms profile determination, which could easily result from one or two
bad spectra.
\begin{table}[tbp]
  \caption{\label{tab:fwhm_balmer}
    Average velocity dispersions in km s$^{-1}$ of the three Balmer
    lines found in this study compared to the values of K00 with
    the rms and average (mean) profiles. The numbers in brackets give the
    $\pm 1\sigma$ confidence interval. A 10\% uncertainty has been
    assumed for our measurements.}
  \begin{center}
    \begin{tabular}{lccc}
      \hline
      \hline
      \rule{0pt}{1.2em}Line&This work&\multicolumn{2}{c}{K00}\\
      &single profile&rms profile&mean profile\\
      \hline\rule{0pt}{1.2em}%
      \rule[-0.6em]{0pt}{1em}\Ha&$4\,424~\left\{_{-442}^{+442}\right.$&
      $1\,152~\left\{_{-\un 76}^{+\un 76}\right.$&
      $2\,434~\left\{_{-164}^{+164}\right.$\\
      \rule[-0.6em]{0pt}{1em}\Hb&$3\,195~\left\{_{-320}^{+320}\right.$&
      $2\,375~\left\{_{-\un 51}^{+\un 51}\right.$&
      $2\,958~\left\{_{-\un 63}^{+\un 63}\right.$\\
      \rule[-0.6em]{0pt}{1em}\Hg&$3\,496~\left\{_{-350}^{+350}\right.$&
      $2\,979~\left\{_{-139}^{+139}\right.$&
      $3\,256~\left\{_{-147}^{+147}\right.$\\
      \hline
    \end{tabular}
  \end{center}
\end{table}

\begin{figure}[tbp]
  \vspace*{10mm}
  \resizebox{\hsize}{!}{\includegraphics{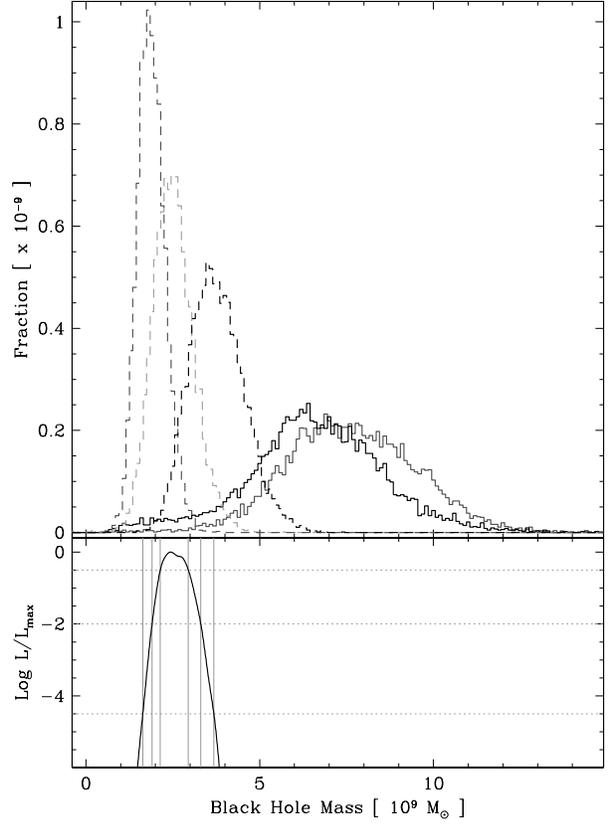}}%
  \caption{\label{fig:massall}
    Top panel: Black-hole mass error distributions for the five
    emission lines. The continuous black and dark grey lines are,
    respectively, the \Lnnv\ and \civ\ histograms as in
    Fig.~\ref{fig:mass}. The dashed lines are \Ha\ (black), \Hb\ (dark
    grey), and \Hg\ (light grey) respectively.  Bottom panel:
    Likelihood function of the mass.  The horizontal grey lines show,
    from top to bottom, the $\pm 1\sigma$, $\pm 2\sigma$, and $\pm
    3\sigma$ levels. The vertical lines show the $\hat{M}_{\pm 1}$,
    $\hat{M}_{\pm 2}$, and $\hat{M}_{\pm 3}$ intervals.}
\end{figure}

The masses obtained for the three Balmer lines are reported in
Table~\ref{tab:mass_balmer} and
\begin{table}[tbp]
  \caption{\label{tab:mass_balmer}
    Average black-hole masses  of \3c\ in units of $10^9\,$M$_\odot$ for
    the three Balmer lines found in this study compared to
    K00 using the rms and average (mean) profiles. The numbers
    in brackets give the $\pm 1\sigma$, $\pm 2\sigma$, and $\pm 3\sigma$
    confidence intervals. The uncertainties on the individual K00 values
    are of the order of 30\%. $\hat{M}$
    is the maximum-likelihood masses obtained from the combinations of
    the three Balmer emission lines and of the two UV and three
    Balmer emission lines, respectively.}
  \begin{center}
    \begin{tabular}{lccc}
      \hline
      \hline
      \rule{0pt}{1.2em}Line&This work&\multicolumn{2}{c}{K00}\\
      &single profile&rms profile&mean profile\\
      \hline\rule{0pt}{1.2em}%
      \rule[-0.6em]{0pt}{1em}\Ha&$3.78~\left\{_{-0.78}^{+0.78}\right.\left\{_{-1.40}^{+1.71}\right.\left\{_{-1.98}^{+2.59}\right.$&
      $0.115$&
      $0.513$\\
      \rule[-0.6em]{0pt}{1em}\Hb&$1.89~\left\{_{-0.39}^{+0.40}\right.\left\{_{-0.73}^{+0.88}\right.\left\{_{-1.06}^{+1.35}\right.$&
      $0.363$&
      $0.564$\\
      \rule[-0.6em]{0pt}{1em}\Hg&$2.51~\left\{_{-0.58}^{+0.58}\right.\left\{_{-1.09}^{+1.30}\right.\left\{_{-2.32}^{+2.04}\right.$&
      $0.459$&
      $0.549$\\
      \hline
      \rule{0pt}{1.2em}%
      \rule[-0.6em]{0pt}{1em}$\hat{M}$ (Balmer)&$2.42~\left\{_{-0.29}^{+0.34}\right.\left\{_{-0.56}^{+0.66}\right.\left\{_{-0.84}^{+1.03}\right.$\\
      \rule[-0.6em]{0pt}{1em}$\hat{M}$ (All)&$2.44~\left\{_{-0.30}^{+0.51}\right.\left\{_{-0.54}^{+0.87}\right.\left\{_{-0.80}^{+1.24}\right.$\\
      \hline
    \end{tabular}
  \end{center}
\end{table}
Figure~\ref{fig:massall} shows the mass error distributions obtained
for the five emission lines. The three Balmer-line mass estimates
agree rather well with each other, but, even though the overlap is not
completely negligible, these estimates seem significantly lower than
the UV ones. The lower mass obtained from the Balmer lines can be due
to the different systematics involved in the UV and optical studies
and, in particular, to the presence of narrow Balmer emission, which
artificially decreases the measured velocity dispersion. A
maximum-likelihood estimate of the mass can be obtained by combining
the \Lnnv, \civ, \Ha, \Hb, and \Hg\ black-hole mass error
distributions. We obtain a mass of $2.44\,10^9$ M$_\odot$. The
$1\sigma$, $2\sigma$, and $3\sigma$ confidence intervals are
$2.14-2.95\,10^9$ M$_\odot$, $1.90-3.31\,10^9$ M$_\odot$, and
$1.64-3.68\,10^9$ M$_\odot$ respectively. We note that the weights of
the Balmer lines overwhelm those of the UV lines.  This is because of
their much narrower uncertainty distributions. However, the UV lines
should be less affected by systematics (the narrow-line emission in
particular).
This maximum-likelihood estimate allows us to set a very secure
$3\sigma$ lower limit on the mass of the black hole of $1.6\,10^9$
M$_\odot$.  Using only the Balmer lines, we find a mass of
$2.42\,10^9$ M$_\odot$, with $1\sigma$, $2\sigma$, and $3\sigma$
confidence intervals of $2.13-2.76\,10^9$ M$_\odot$, $1.86-3.08\,10^9$
M$_\odot$, and $1.58-3.45\,10^9$ M$_\odot$, respectively.

\section{Discussion}
\label{sec:disc}

\subsection{Black-hole mass and Eddington ratio}
The main result of this study is the discrepancy between the masses
found by K00 and our estimates, since they differ by a factor ranging
from 12 up to 60. As our study of the Balmer lines shows,
discrepancies affect both the estimates of the delays and of the line
widths. A general argument in favor of our mass estimate is the good
qualitative agreement between the individual estimates for the \Lya,
\civ, and the three Balmer lines. The largest difference between the
UV and Balmer lines is a factor three, while the systematics in the
two sets of data can be quite different. By contrast, the K00
estimates, reported in Table~\ref{tab:mass_balmer}, differ by a factor
5 between each other, and about $4.5$ for \Ha\ only, although all the
estimates have been obtained with a consistent data set. In our study,
we could not determine the rms profiles of the Balmer lines (not even
their average profiles), because of the lack of data.  Therefore, it
is quite probable that a large part of the UV-Balmer discrepancy would
be alleviated if the same methodology could be applied to the Balmer
lines as well. It must be noted that all Balmer profiles are affected
by contaminating emission lines, in addition to the narrow-line Balmer
emission: [N{\sc ii}] $\lambda$6548 and [N{\sc ii}] $\lambda$6583 for
\Ha; [O{\sc iii}] $\lambda$4959 and [O{\sc iii}] $\lambda$5007 for
\Hb; [O{\sc iii}] $\lambda$4363 for \Hg. The [O{\sc iii}] lines being
particularly strong in the wing of \Hb, its red profile had to be
excluded from our analysis. The \Hb\ region is moreover seriously
contaminated by Fe{\sc ii} pseudo-continuum emission
\citep{BoroGree-1992-EmiLin}, making the line profile of \Hb\ quite
uncertain. The remaining narrow lines, plus the narrow-line Balmer
emission, may significantly modify the true broad-line profiles.  As
they fall close to the Balmer line centers, their effect is to reduce
the line second moments, providing underestimated black-hole masses.
We thus conclude that, although affected by larger statistical errors,
the UV-line estimates very probably suffer from smaller systematic
errors.  Therefore, the best mass estimate is the combined UV mass:
$6.59\,10^9$\, M$_\odot$, with a $3\sigma$ lower limit of
$3.44\,10^9$\, M$_\odot$. Including the Balmer lines, the $3\sigma$
lower limit is found to be $1.6\,10^9$ M$_\odot$. This lower limit is
very robus, and should only increase if the systematic errors are
reduced.

We point out that this fairly high mass estimate is obtained under the
hypothesis that the gas motion is virialized and (close to) isotropic.
As \3c\ has a superluminal jet \citep[e.g.,][]{Cour-1998-BriQua},
\3c's accretion disk is very probably close to face-on (assuming the
disk is perpendicular to the jet), so projected velocities of
particles in the disk would be greatly reduced by the geometry. If the
emission lines originate from the disk, the mass estimate should be
multiplied by a factor $1/\sin{i}>6$, if the angle between the
perpendicular to the disk and the line of sight is $i<10^\circ$.  On
the basis of the response of the UV lines, we also concluded in \PI\ 
that accretion-disk geometries of the broad line region can be
excluded in this object.

Our comparison between the red and full profiles for the \Lya\ and
\civ\ emission lines shows that inclusion of the possibly
outflowing gas seen in \PI\ would misleadingly increase the estimate
of the mass of the black hole by approximately 15--30\%. This is far
from negligible, but it also shows that the order of magnitude of our
result is not affected by the choice of using only the red profile.

From \citet{Cour-1998-BriQua} and \citet{TuerEtal-1999-ThiYea}, we
estimate the bolometric luminosity of \3c\ to be approximately
$10^{47}$ erg\,s$^{-1}$ (using
$H_0\!=\!70$\,km\,s$^{-1}$\,Mpc$^{-1}$), giving an Eddington mass
of about $10^{9}$\,M$_\odot$. Using the 5100\,\AA\ flux as an
estimate of the bolometric luminosity and a different cosmology, K00
find a comparable value of $\sim 0.5\,10^{9}$\,M$_\odot$. According to
the results of K00, the ratio between the mass derived from the
Eddington luminosity and that from the reverberation method is
therefore in the range 1--4, depending on which mass estimates are
used, which makes \3c\ a very strong candidate for super-Eddington
accretion. Using our new reverberation-method mass, the Eddington
ratio is estimated to be about $0.15$ and, using the individual
estimates, falls in the range 0.13--0.53, i.e.\ in a fully
sub-Eddington regime.

\subsection{Size of the broad-line region}
The size of the Balmer-emitting broad-line region in AGN has been
investigated by several authors
\citep[][K00]{KoraGask-1991-RadLum,WandEtal-1999-CenMas}. While the
first two studies find a BLR size $R_{\mathrm{BLR}}$ scaling roughly
as $L^{0.5}$, K00 find a steeper relationship $R_{\mathrm{BLR}}\sim
L^{0.700\pm 0.033}$ using a sample of 34 objects that were
investigated using the reverberation method. The $L^{0.5}$
relationship is expected if gas density and ionization parameters are
independent of the source luminosity in the emitting region.  However,
\citet{Vest-2002-DetCen} correctly points out that the K00
relationship is calculated with a method unable to take the strong
intrinsic scatter into account, which produces biased slopes and too
low uncertainties.  Applying the BCES method of
\citet{AkriBers-1996-LinReg} on the K00 data and using a modified BLR
size for \object{NGC\,4051} from \citet{PeteEtal-2000-XraOpt}, she
found $R_{\mathrm{BLR}}\sim L^{0.58\pm 0.09}$, perfectly compatible
with $L^{0.5}$. We can refine this relationship by using the new
$R_{\mathrm{BLR}}$ for \3c found in this work and by including new
measurements for \object{NGC\,3783} \citep{OnkePete-2002-MasCen}, and
\object{NGC\,3227} \citep{OnkeEtal-2003-BlaHol}. \3c\ has an important
weight in the relationship, because it is the highest-luminosity
object in the sample. From a maximum-likelihood estimator on the lag
of the three Balmer lines, we have $R_{\mathrm{BLR}}=986^{+21}_{-37}$
days.  Applying the BCES estimator, we obtain
\begin{equation}
R_{\mathrm{BLR}}\sim\lambda L_\lambda(5100\,\AA)
^{0.59\pm 0.09}.
\end{equation}
The two relationships are perfectly compatible with each other and
still compatible with a $L^{0.5}$ relationship. Excluding NGC\,4051 as
proposed in \citet{Vest-2002-DetCen}, the slope becomes $0.67\pm
0.10$, but it is not clear why this object should be discarded, as its
distance to the BCES estimator does not appear unreasonable,
considering its uncertainty. It must be noted, however, that this object
has a very strong weight in the relationship.

\citet{McluJarv-2002-MeaBla} recalculated the $L_\lambda(5100\,\AA)$
luminosities using fluxes from \citet{NeugEtal-1987-ConEne} and a
modern cosmology ($\Omega_m\!=\!0.3$, $\Omega_\Lambda\!=\!0.7$,
$H_0\!=\!70$\,km\,s$^{-1}$\,Mpc$^{-1}$), and found a slope flatter
than K00. We find here that, while the slope of the bisector is
slightly lower, it is not significantly affected by the new
luminosities, showing that the K00 luminosities are sufficiently
accurate.

\subsection{Mass-luminosity relationship}
With the same sample, K00 investigated the mass-luminosity
relationship in 34 Seyfert 1 and QSOs.  They found that the
relationship deviates significantly from the $L\sim M$ relationship,
which is expected if the Eddington ratio
$L_{\mathrm{bol}}/L_{\mathrm{Eddington}}$ is independent of the
black-hole mass. They find $M_{\mathrm{BH}}\sim L^{0.5\pm 0.1}$,
meaning that Seyfert 1 galaxies have low accretion rates compared to
QSOs. It even strongly suggests that very luminous QSOs must accrete
in the super-Eddington regime. This result has indeed sparked a lot of
interest in the theoretical investigation of super-Eddington models
\citep{CollEtal-2002-QuaAcc}.

\begin{figure}
  \resizebox{\hsize}{!}{\includegraphics{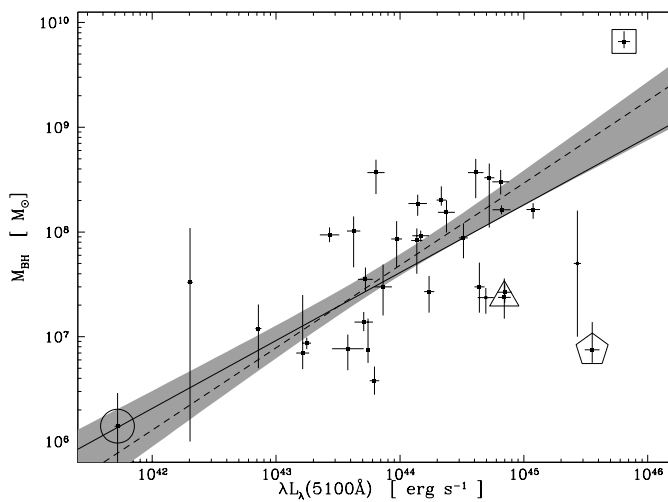}}%
  \caption{\label{fig:Mbh}
    Mass-luminosity for the 34 objects in the K00 sample. The solid
    line is the BCES bisector. The dashed line is the BCES bisector
    obtained after discarding unreliable objects. The shaded area
    shows the 1$\sigma$ error domain. The square, circle, triangle and
    pentagon respectively show \3c, NGC~4051, PG\,1613+658, and
    PG\,1704+608.}
\end{figure}

As we find that \3c\ has a lower accretion rate than previously thought, we
reexplore the mass-luminosity relationship using the BCES bisector.
With the new \3c\ mass found here and the updated masses of the three
Seyfert galaxies, we find
\begin{equation}
M_{\mathrm{BH}}\sim \lambda L_\lambda(5100\,\AA)^{0.65\pm 0.18}.
\end{equation}
Figure~\ref{fig:Mbh} shows the updated mass-luminosity relationship.

We caution, however, that this relationship can be disturbed by a few
ill-measured masses. For instance, in the K00 sample two objects,
\object{PG\,1613+658} and \object{PG\,1704+608}, have masses differing
by a factor 5 or larger, depending on whether the rms- or mean-profile
method is used. If we discard these objects as too unreliable, we
obtain
\begin{equation}
M_{\mathrm{BH}}\sim \lambda L_\lambda(5100\,\AA)^{0.79\pm 0.19}.
\end{equation}

The reverberation-mass sample is therefore compatible with an
Eddington ratio independent of the black-hole mass.  This is
consistent with the conclusion of \citet{WooUrry-2002-ActGal}, who
investigated the mass-luminosity relationship in a sample of 377 AGN
whose masses had been estimated using different methods, including
reverberation. If the absence of relationship is confirmed, it means
that there may currently be no need for super-Eddington-luminosity
models.

Here as well, using the luminosities from \citet{McluJarv-2002-MeaBla}
does not significantly affect the bisector estimates.

\section{Conclusion}
\label{sec:conc}
The most important consequence of this work is to show that the
reverberation method can provide quite different results on a single
object, as we find a black hole mass of about $6.59\,10^9$\,M$_\odot$,
an order of magnitude larger than in the previous study of K00. This
result is considerably stronger than that of \citet{Krol-2001-SysErr},
because it shows that serious discrepancies can result from the
measurement of the delays and line widths and not only from the
underlying assumptions of the method. In this study, we benefited
from a much better determination of the delay between the lines and
the continuum, thanks to the use of the ultraviolet continuum with a
better sampling.  However a large fraction of the discrepancy is caused
by the determination of the line widths, for which we can propose only
tentative explanations.

The mass determined from Balmer lines is about a factor 3 smaller than
that obtained from the UV lines, but because of the lack of adequate
data we could not determine the rms profile of the Balmer lines and
had to rely on a less optimal velocity-dispersion measurement method.
Thus at least part of the discrepancy can be explained by the
different systematics and in particular by the inclusion of
narrow-line emission in the line profile. Our Balmer-line mass is, in
any case, at least a factor 5 larger than the previous study's
estimate. We are able to put a very robust $3\sigma$ lower limit on
the mass of the black hole in \3c\ of $1.6\,10^9$\,M$_\odot$.

We studied the mass-luminosity relationship of supermassive black
holes determined with the reverberation method. Using our new \3c\ 
mass estimat, and updated estimates for three Seyfert galaxies, we
find that the case for a correlation between the Eddington ratio and
the black-hole mass in the reverberation-method sample is far from
established.

We add finally that, in a very recent in-depth reanalysis of the
exhaustive data set from the AGN Watch collaboration
\citep{AlloEtal-1994-IntAgn}, \citet{PeteEtal-2004-CenMas} obtained a
new estimate of the mass of \3c\ of $0.89\pm0.19\,10^9$\,M$_\odot$.
Using our delay estimates for the Balmer lines, the estimated mass
would be approximately $2.5\,10^9$\,M$_\odot$, perfectly compatible
with our own estimate using the Balmer lines
($2.42\,10^9$\,M$_\odot$). \citet{PeteEtal-2004-CenMas} also give a
much improved estimate of the slope of the mass-luminosity
relationship $M_{\mathrm{BH}}\sim L^{0.79\pm 0.09}$, again in complete
agreement with our work.

\acknowledgements{SP acknowledges a grant from the Swiss National Science Foundation.}

\bibliographystyle{apj}
\bibliography{biblio}

\end{document}